\documentclass[conference]{IEEEtran}
\IEEEoverridecommandlockouts

\usepackage{setspace}
\usepackage{cite}
\usepackage[utf8]{inputenc}
\usepackage{siunitx} 
\usepackage{epsf}\epsfverbosetrue
\usepackage{graphics,epsfig,color}
\usepackage{graphicx,epsfig}
\usepackage{epsfig}
\usepackage{multirow}
\usepackage{xcolor}
\usepackage{subfigure}
\usepackage{caption}
\usepackage{footnote}
\usepackage{booktabs}
\usepackage{threeparttable}
\usepackage{color}
\usepackage{float}
\usepackage{url}
\usepackage{graphicx}
\usepackage{verbatim} 
\usepackage{footnote} 
\usepackage{latexsym} 
\usepackage{amsmath}
\usepackage{amssymb} 
\usepackage{sidecap} 
\usepackage{color}
\usepackage{wrapfig}
\usepackage{algorithm}
\usepackage{eso-pic}
\usepackage{fix-cm}
\usepackage{mcite}
\usepackage{layout}
\usepackage{amsmath}
\usepackage{textcomp}
\usepackage{multirow}
\usepackage{soul}
\usepackage{amsmath}
\usepackage{verbatim}
\usepackage{bbding}
\usepackage{tabularx} 
\usepackage{comment}
\usepackage{algorithm}

\usepackage{algpseudocode}
\usepackage{multirow}
\usepackage{booktabs} 
\usepackage{lipsum} 

\begin{document}

\title{End-to-End Waveform and Beamforming Optimization for RF Wireless Power Transfer\\
\thanks{This work is partially supported in Finland by the Finnish Foundation for Technology Promotion, the Research Council of Finland (Grants 348515 and 346208 (6G Flagship)) and by the European Commission through the Horizon Europe/JU SNS project Hexa-X-II (Grant Agreement no. 101095759)}
}
\author{
 {Abdul Basit Khattak$^{}$, Onel L. A. L\'{o}pez$^{}$, Amirhossein Azarbahram$^{}$, Deepak Kumar$^{}$, Matti Latva-aho$^{}$}
\vspace{1mm} \\

$^{}$\normalsize Centre for Wireless Communications (CWC), University of Oulu, Finland\\

Emails: \normalsize{\{abdul.khattak, onel.alcarazlopez, amirhossein.azarbahram, deepak.kumar,  matti.latva-aho\}@oulu.fi} \\
\vspace{-6mm}
}

\maketitle

\begin{abstract}
Radio frequency (RF) wireless power transfer (WPT) is a key technology for future low-power wireless systems. However, the inherently low end-to-end power transfer efficiency (PTE) is challenging for practical applications. The main factors contributing to it are the channel losses, transceivers' power consumption, and losses related, e.g., to the digital-to-analog converter (DAC), high-power amplifier, and rectenna. Optimizing PTE requires careful consideration of these factors, motivating the current work. Herein, we consider an analog multi-antenna power transmitter that aims to charge a single energy harvester. We first provide a mathematical framework to calculate the harvested power from multi-tone signal transmissions and the system power consumption. Then, we formulate the joint waveform and analog beamforming design problem to minimize power consumption and meet the charging requirements. Finally, we propose an optimization approach relying on swarm intelligence to solve the specified problem. Simulation results quantify the power consumption reduction as the DAC, phase shifters resolution, and antenna length are increased, while it is seen that increasing system frequency results in higher power consumption. 
\end{abstract}
\vspace{-1mm}
\begin{IEEEkeywords}
Radio frequency wireless power transfer, power transfer efficiency, waveform optimization, energy beamforming.
\end{IEEEkeywords}
\vspace{-0.8em}
\section{Introduction}
Wired charging and battery replacements hinder sustainable connectivity in industrial Internet of Things (IoT) installations. Moreover, battery-powered solutions for IoT devices face limitations due to short battery lives and replacement challenges, which are neither cost-effective nor environmentally friendly due to their associated waste. Thus, the research community has focused on energy harvesting (EH) methods as a promising solution to recharge batteries externally, which prevents maintenance and replacements \cite{lopez2021massive, lopez2023high}.\par{} 
EH sources can be categorized into ambient, i.e., energy readily available in the environment, and dedicated, i.e., deliberate energy transmissions. The latter category is facilitated by wireless power transfer (WPT) technologies, e.g., inductive coupling, magnetic resonance coupling, laser power beaming, and radiative radio frequency (RF). In contrast to other methods, RF-WPT offers inherent advantages like compact energy receiver (ER) designs, multiuser support, and greater charging radii \cite{lopez2024zero}. Note that RF-EH/WPT are essential enablers of backscattering systems, providing them with the necessary power autonomy, extended range, and environmental sustainability required for diverse IoT and RFID applications. However, its low power transfer efficiency (PTE) is a critical challenge that needs to be carefully addressed \cite{lopez2021massive, song2021advances}.\par{} 
Recent research has focused on waveform design, beamforming optimization, and link-level assessments to improve the efficiency of RF-WPT systems, e.g., \cite{zeng2017communications, moghadam2017waveform, azarbahram2023energy}. Notice that proper waveform can enhance the PTE by leveraging the non-linear behavior of the system's key components, e.g., rectenna and high power amplifier (HPA). Specifically, it is shown that utilizing a high peak-to-average power ratio signal can improve the rectenna's RF-to-direct current (DC) conversion efficiency \cite{clerckx2018fundamentals}. Meanwhile, energy beamforming (EB) can be used to focus the RF signal toward the ER to compensate for the channel losses, thereby, enhancing the amount of available RF power for EH purposes \cite{azarbahram2023energy}. Note that EB flexibility and potential gains are determined by the transmitter architecture, which significantly impacts the power consumption of RF-WPT systems \cite{rosabal2023sustainable}. In a fully-digital architecture, each transmitting antenna requires a dedicated RF chain, with its corresponding HPA. Moreover, the HPA's signal amplification requires a DC power source, which accounts for the majority of the system's power consumption. Notably, the HPA introduces non-linear signal distortion that requires precise modeling \cite{joung2014survey}. The digital-to-analog converter (DAC) is another crucial component with its cost and power consumption scaling with the resolution and sampling rate \cite{cui2005energy}. Overall, the high complexity and cost associated with fully-digital structures make it impractical for applications requiring massive implementations \cite{kalcher2018fully}. Meanwhile, the analog counterpart is a simpler architecture as it can operate with a single RF chain at the cost of single beam transmissions and less spatial flexibility. Moreover, the analog beamforming reduces the power transmitter (PT) energy consumption, hardware complexity, and the overall deployment cost \cite{ahmed2018survey}.\par{} 
In this paper, we consider a multi-antenna analog RF-WPT system comprising multiple phase shifters (PSs) with a limited resolution to promote low-cost implementations. Also, we consider a near-field wireless channel to emphasize the role of near-field WPT systems \cite{lopez2023high}. Our contributions are:
\begin{figure*}[t!]
    \centering
\includegraphics[width=0.8\textwidth]{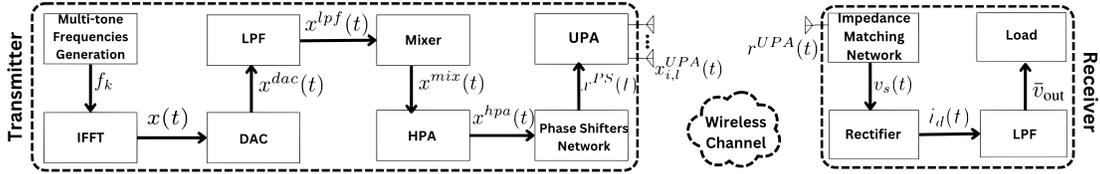}
    \caption{Block diagram of the proposed RF-WPT system.}
    \label{fig_1}
    \vspace{-1.6em}
\end{figure*}
\begin{itemize}
\item We consider the main non-linear components of the system, e.g., HPA and rectenna. Although the authors in \cite{clerckx2016waveform} consider these non-linearities, our EH model is a generic model \cite{moghadam2017waveform} based on circuit analysis that accurately captures the EH nonlinearity without relying on Taylor approximation as in \cite{clerckx2016waveform, shen2021joint}. Additionally, we derive the closed-form expression for the DC output voltage. 
\item  We formulate a joint waveform and beamforming design problem aiming to minimize a single-ER system's power consumption and meet the EH requirements. This differs from \cite{clerckx2016waveform, zhang2023waveform, feng2022waveform}, which focused on increasing the harvested power given a transmit power. Additionally, fully-digitial and hybrid architectures are considered in \cite{clerckx2016waveform, azarbahram2023waveform}, but herein, we consider an analog architecture with limited resolution. Note that the complexities associated with the HPA  modeling, the EH model, and the limited-resolution phase shifts make the existing optimization frameworks for RF-WPT systems inapplicable to our system, which calls for novel solutions.
\item  We propose an optimization approach relying on particle swarm optimization (PSO) to solve the joint waveform and beamforming design problem. This is done by incorporating the user’s harvesting DC power requirements in the objective function of the optimization problem, thus compelling the algorithm to discover a feasible solution before minimizing the power consumption. Hereby, we cope with the non-convex nature and extensive non-linearities of the problem, introduced by the HPA input-output relation, the HPA power consumption model, the phase model of the low-resolution PS, and the EH model. 
\item  We show numerically that the overall power consumption decreases as the DAC/PSs resolution, and antenna length increase, while it increases with the operation frequency.
\end{itemize}
\textbf{Structure}: Section II introduces the system model, Section III presents the harvested power and the power consumption, and the problem formulation together with the optimization framework is discussed in Section IV. Numerical analysis is provided in Section V, while Section VI concludes the paper.
\vspace{-0.8em}
\section{System Model} 
\vspace{-0.4em}
We consider an RF-WPT system comprising three key building blocks, i.e., a PT, wireless channel, and ER, as illustrated in Fig.~\ref{fig_1}. The PT utilizes a uniform planar array (UPA) with half-wavelength spaced radiating elements in the horizontal and vertical directions, represented as $H$ and $V$, respectively. The total number of antenna elements is given by $N=V \times H$. We consider an analog architecture where the output of a single RF chain is connected to all the radiating elements, which are equipped with PSs. 
\vspace{-3mm}
\subsection{Transmitter}
\vspace{-2mm}
\subsubsection{Multi-tone Signal Generation}
Recall that utilizing a multi-tone signal can improve the end-to-end PTE by exploiting non-linearities at the transmitter and receiver \cite{zeng2017communications, clerckx2018fundamentals}. Hence, we consider a signal with $K$ tones and frequency spacing $\Delta_f$ between adjacent tones. Thus, the baseband signal bandwidth and baseband frequency of the $k$th tone is $B W = K \Delta_f$ and $f_k=k \Delta_f$, respectively. An inverse discrete Fourier transform (IDFT) block converts the signal to the time domain. With the sampling frequency $\tilde{f}_s$ and the sampling time $ T_s=1 / \tilde{f}_s$, the IDFT for a signal $x(t)$ can be expressed as
\vspace{-0.3em}
\begin{equation}
\label{eq:1}
x\left(n T_s\right)=\frac{1}{K} \sum_{k=0}^{K-1} {X_{k}} e^{j 2 \pi f_k n T_s + \phi_{k}}, n=0, 1, 2, …, \vspace{-0.3em}
\end{equation}
where ${X_{k}}$ and ${\phi_{k}}$ are the amplitude and phase of the $k$th tone.  
\vspace{-0.5em}
\subsubsection{DAC}
We assume a DAC with a resolution of $n_b$ bits and input signal's voltage range $[-A, A]$. Thus, based on \eqref{eq:1}, we can write $\sum_{k=0}^{K-1} |X_k| \le A $, while we assume a linear quantization with step size $\Delta = \frac{2A}{2^{n_b}}$. Hereby, $x(t)$ is quantized by rounding it to the nearest value of the quantization step size $\Delta$, i.e., $x^{dac}(t)=\left\lfloor x(t)/\Delta\right\rceil \Delta$. 
\subsubsection{Low-pass Filter (LPF)}
 An LPF filters out undesired high frequencies introduced by DAC. The discrete
Fourier transform (DFT) of the quantized signal, i.e., ${X}_{}^{dac}=\operatorname{DFT}\{x^{dac}(t)\}$, is considered for spectrum analysis, while the LPF output is ${X}^{lpf} = \left. \ {X}^{dac} \right|_{f\leq B W}$.
\subsubsection{Mixer}
The mixer upconverts the signal to the desired RF, i.e., $f_{ca}$, by a local oscillator with its output given by 
\begin{equation}
x^{mix}(t) = \Re\left\{x^{lpf}(t) e^{j 2 \pi f_{c a} t}\right\},
\end{equation}
where $x^{lpf}(t)$ is the analog signal at the output of the LPF and $\Re\{.\}$ is the real operator.
\subsubsection{HPA}
We employ the Rapp model for solid-state power amplifiers \cite{rapp1991effects} to capture the HPA's non-linear behavior. Hereby, the output signal of the HPA is given by
\vspace{1mm}
\begin{equation}
\label{eq:11}
x^{hpa}(t) ={Gx^{mix}(t)}{\left(1 + \left(G \lvert x^{mix}(t) \rvert/A_s\right)^{2 \beta}\right)^{-\frac{1}{2 \beta}}},
\end{equation}
\noindent where $G$ represents the amplifier gain, $A_s$ is the saturation voltage, while $\beta$ is the smoothening parameter of the amplifier.
\vspace{-0.4em}
\subsubsection{Transmit Signal}
The RF signal goes through PSs with a finite resolution of $B$ bits for phase adjustment before being fed into the UPA elements. Hence, the signal transmitted by the $l^{t h}$ element in the $i^{t h}$ row of the UPA is given by
\begin{equation}
\label{eq:12}
x_{i, l}^{UPA}(t) = ({1}/{\sqrt{L_{ps} N}}) e^{-i \frac{2 \pi b_{i,l}}{2^B}} x^{hpa}(t),
\end{equation}
where $\text{} b_{i,l}\in \{0,1, \ldots, 2^B-1\}$ is the corresponding quantized phase shift and $L_{ps}$ represents the insertion loss.
\vspace{-1.25mm}
\subsection{Channel Model}
\vspace{-1.2mm}

We consider a near-field channel model \cite{zhang2022beam}, which also applies to far-field conditions. Let us proceed by defining the Cartesian coordinates of the $l$th antenna element in the $i$th row as $\mathbf{p}_{i,l}=(x_{i,l}, y_{i,l}, z_{i,l})$. Then, the channel coefficient between the ER, located at $\mathbf{p}=\left(x, y, z\right)$, and the $l$th element of the $i$th row of the UPA corresponding to the $k$th tone is given as
\textbf{}
\begin{equation}
\label{eq:13}
H_{i,l,k}= A_{i, l, k} e^{\frac{-j 2 \pi}{\lambda_k}\left||\mathbf{p}-\mathbf{p}_{i, l}\right||_2}.
\end{equation}
Herein, $\frac{ 2 \pi}{\lambda_k}\left||\mathbf{p}-\mathbf{p}_{i, l}\right||_2$ is the phase shift caused by the distance traveled by the $k$th tone from the transmitter to the receiver and $\lambda_k$ is the wavelength of the $k$th tone.
The corresponding channel gain coefficient is given by \cite{ellingson2021path}
\textbf{}
\begin{equation}
\label{eq:14}
A_{i, l, k}=\lambda_k\sqrt{F\left(\Theta_{i, l}\right)}/({4 \pi\left||\mathbf{p}-\mathbf{p}_{i, l}\right||_2}),
\end{equation}
where $\Theta_{i, l}=\left(\theta_{i, l}, \psi_{i, l}\right)$ represents the elevation-azimuth pair from the $l$th element of the $i$th row of the UPA to the ER. The radiation profile $F\left(\Theta_{i, l}\right)$ of each element is modeled as \cite{ellingson2021path}
\textbf{}
\begin{equation}
\label{eq:15}
F\left(\Theta_{i, l}\right)= \begin{cases}2(b+1) \cos ^b\left(\theta_{i, l}\right), & \theta_{i, l} \in[0, \pi / 2], \\ 0, & \text { otherwise, }\end{cases}
\end{equation}
where $b$ is the boresight gain. The RF signal at the ER is 
\textbf{}
\begin{equation}
\label{eq:16}
r^{UPA}(t)=\sum_{i=1}^{V} \sum_{l=1}^{H} \sum_{k=0}^{K-1} H_{i, l, k} x_{i, l}^{UPA}(t).
\end{equation}
\vspace{-1.4em}
\subsection{Energy Receiver}
\begin{figure}
    \centering
    \includegraphics[width=0.7\columnwidth]{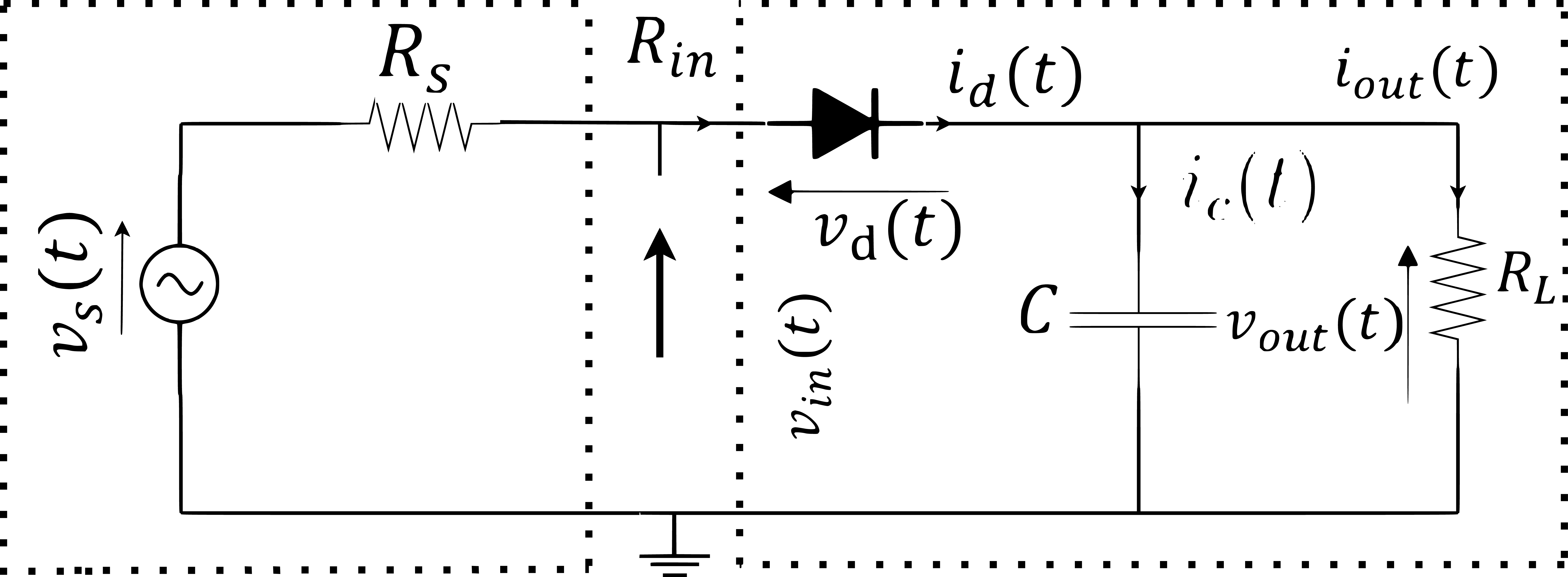}
    \vspace{-1mm}
    \caption{Antenna equivalent circuit (left) and the rectifier (right).}
    \label{fig_2}
    \vspace{-4mm}
\end{figure}

\vspace{-0.2em}
The ER structure, shown in Fig.~\ref{fig_2} \cite{moghadam2017waveform}, comprises i) an antenna, equivalently represented as a voltage source connected in series with a resistance, $R_s$, with $v_s(t)=2 \sqrt{R_s} r^{UPA}(t)$ being the source voltage; ii) an impedance matching network matching the antenna impedance to the impedance of the rectifier circuitry for maximum power delivery; iii) and a rectifier converting the RF alternating current (AC) signal received from the antenna into a DC signal used to charge the load. We assume perfect impedance matching, i.e.,  $R_s = R_{\mathrm{in}}$, where $R_{\mathrm{in}}$ refers to the input resistance of the rectifier. The rectifier includes a single diode and an LPF connected to a load resistance $R_L$. The LPF is used to filter out any remaining RF signals and noise from the rectified DC output. \par{}
The input voltage of the rectenna is given by $v_{\mathrm{in}}(t) = \frac{v_s(t)}{2} = \sqrt{R_s} r^{UPA}(t)$, while output voltage is $v_{\text {out }}(t)=\bar{v}_{\text {out }}+\tilde{v}_{\text {out }}(t)$, where $\bar{v}_{\text {out }}$ represents the DC component and $\tilde{v}_{\text {out }}(t)$ represents the AC component. One can obtain \cite{moghadam2017waveform}
\textbf{}
\begin{equation}
\label{eq:21}
 e^{\frac{\bar{v}_{\text {out}}}{\eta_{} V_0}}\left(1+\frac{\bar{v}_{\text {out}}}{R_L I_0}\right)=\frac{1}{T} \int_T e^{\frac{\sqrt{R_s} r^{UPA}(t)}{\eta_{} V_0}} dt,
\end{equation}
where $I_0$ represents the reverse bias saturation current of the diode, $V_0$ is its thermal voltage, and $\eta_{}$ is the ideality factor. Moreover, $T$ represents the integrating period. Finally, the DC power delivered to the load is
\textbf{}
\begin{equation}
\label{eq:33}
\bar{p}_{\text {out}}={\bar{v}_{\text {out}}^2}/{R_L}.
\end{equation}
We omit some details here due to space limitations and one can follow \cite{moghadam2017waveform} for more details.
\vspace{-3mm}
\section{Harvested Power \& Power Consumption}
\vspace{-1mm}
In this section, we derive the closed-form expression for DC output voltage and formulate the system power consumption. 
\vspace{-2mm}
\subsection{DC Output Voltage}
\vspace{-1mm}
The relationship between the input and output voltages of the rectifier circuit is captured in \eqref{eq:21}. To proceed further, we rewrite \eqref{eq:21} by multiplying its both sides by $\frac{R_L I_0}{\eta V_0}e^{\frac{R_L I_0}{\eta V_0}}$ as
\begin{align}
    &e^{\left(\frac{\bar{v}_{\text{out}} \!+\! R_L I_0}{\eta V_0}\right)} \left(\frac{R_L I_0 \!+\! \bar{v}_{\text{out}}}{\eta V_0}\right) =\frac{ e^{\frac{R_L I_0}{\eta V_0}}R_L I_0}{\eta V_0 T} \int_T e^{\frac{\sqrt{R_s} r^{\text{UPA}}(t)}{\eta V_0}} dt  \notag \\
    &\frac{R_L I_0 + \bar{v}_{\text{out}}}{\eta V_0} \notag 
    = \text{W}\left(e^{\frac{R_L I_0}{\eta V_0}}\frac{R_L I_0}{\eta V_0 T} \int_{T} e^{\frac{\sqrt{R_s} r^{\text{UPA}}(t)}{\eta V_0} }dt   \right) \notag \\ 
    &\bar{v}_{\text{out}} \!=\! \eta V_0 \text{W}\left( e^{\frac{R_L I_0}{\eta V_0}}\frac{R_L I_0}{\eta V_0 T} \int_{T} e^{\frac{\sqrt{R_s} r^{\text{UPA}}(t)}{\eta V_0} } dt\right) \!-\! R_L I_0,
    \label{eq:single}
\end{align}
where the last line comes from using the product log or Lambert W function $\text{W}[\cdot]$. Moreover, the integral in \eqref{eq:single} is evaluated by taking the mean of the time domain samples of exponent. Then, $\bar{p}_{\text {out}}$ can be obtained using
\eqref{eq:33}.
\vspace{-2mm}
\subsection{Power Consumption}
\vspace{-1mm}
The overall power consumption of the system is the sum of the power consumed by each component at the transmitter side. Thus, the total power consumption of the system is
\textbf{}
\begin{equation}
\label{eq:40}
P_c=P_{dac}(n_b, \tilde{f}_s)+P_{mix}+P_{lo}+P_{hpa}+P_{s},
\end{equation}
 where $P_{dac}(n_b, \tilde{f}_s)$ is the power consumed by the $n_b$-bit DAC functioning with sampling frequency $\tilde{f}_s$, while $P_{hpa}$, $P_{mix}$, and $P_{lo}$ are the power consumed by the HPA, mixer, and local oscillator, respectively. Moreover,  $P_{s}=\frac{1}{K} \sum_{k=0}^{K-1} \left| X_k \right|^2$ is the power of the generated signal.
\subsubsection{DAC Power Consumption}
The power consumed by the DAC can be written as \cite{cui2005energy}
\textbf{}
\begin{equation}
\label{eq:35}
P_{dac}(n_b, \tilde{f}_s) \approx \alpha\big[V_{d d} I\left(2^{n_b}-1\right)+C_p \tilde{f}_s V_{d d}^2 n_b\big]/2,
\end{equation}
where $V_{d d}$ is the voltage of the power supply, $I$ is the unit current source that corresponds to the least significant bit, and $C_p$ is the parasitic capacitance of the switches used to choose the DAC's supported states, while the second-order effects are captured by the correction factor $\alpha$.
\subsubsection{HPA Power Consumption}
 The input and output power of HPA is given by $P_{i n}={\mathbb{E}(\left|x^{mix}(t)\right|^2)} / {R_{in}}$ and $P_{\text {out }}={\mathbb{E}\left(|x^{hpa}(t)\right|^2)} / {R_{out}}$, respectively, where $R_{in}$ and $R_{out}$ are the input and output resistance of the HPA. The power consumption of the HPA can then be expressed as $P_{hpa}={P_{o u t}-P_{i n}}$.
 \vspace{-4mm}
\section{Optimization Framework}
\vspace{-1mm}
We focus on the joint waveform and beamforming optimization to minimize the power consumption while satisfying the user's DC power requirement, specified by ${P}_{dc}$, i.e., \vspace{-1mm}
\begin{subequations}
\label{eq:42}
\begin{align}
    & \underset{{X_{k}}, \phi_{k}, b_{i,l}}{\text{minimize}} \quad P_c \label{eq:1a} \\
    & \text{subject to} \quad \bar{p}_{\text {out}} \geq {P}_{dc}. \quad  \label{eq:1b}
\end{align}
\end{subequations}
 Note that both \eqref{eq:1a} and \eqref{eq:1b} are non-convex functions of the optimization variables and deal with extensive non-linearities caused by HPA and rectenna.
 \vspace{-2mm}
\subsection{Waveform and Beamforming Optimization}
\vspace{-1mm}
It is challenging to solve problem \eqref{eq:42} directly; thus, we reformulate it as \vspace{-1mm}
\begin{equation}
\label{eq:44}
\underset{{X_{k}}, \phi_{k}, b_{i,l}}{\text{minimize}} \quad P_c+ \Upsilon\left({P}_{dc}-\bar{p}_{\text{out}}\right),
\end{equation}
which is an unconstrained optimization problem. Moreover, $\Upsilon(x)$ in \eqref{eq:44} is an indicator function being $\infty$ $\text{if } x \geq 0$ and $0$ $\text{otherwise}$. This indicator function facilitates removing the constraint \eqref{eq:1b} and transforming \eqref{eq:42} into \eqref{eq:44}. Hereby, the optimization problem is compelled to discover feasible solutions, leading the second term to become zero, prior to minimizing the power consumption of the system. Note that problem \eqref{eq:44} is highly non-linear and non-convex, which can be solved using meta-heuristic algorithms, such as PSO.\par{}
PSO is inspired by the social behavior of swarms in nature. It involves a swarm of particles $N_p$, each representing a potential solution, moving in the solution space. These particles adjust their positions based on their own best experiences $\mathbf{p}_{b_i}$ and those of their neighbors $\mathbf{g}_b$, leading to convergence towards a solution after a sufficient number of iterations \cite{kennedy1995particle}. Moreover, $\mathbf{p}_{b_i}$ is defined as the best position that the $i$th particle has found during the optimization procedure. Additionally, $\mathbf{g}_b=\xi\left(\mathbf{z}_{i^*}\right)$ is the best objective value in the entire swarm, where $\xi$ is the objective function of the optimization problem. \par{}
Let us proceed by defining $\mathcal{S}_{}=\{\mathbf{z}_{1}, \mathbf{z}_{2}, \ldots, \mathbf{z}_{N_p}\}$ as the swarm set for the optimization problem, where $\mathbf{z}_{i}=\big[{{X}}_{i, 1}^T, \ldots, {{X}}_{i, K}^T, \boldsymbol{\phi}_{i, 1}^T, \ldots, \boldsymbol{\phi}_{i, K}^T, \bar{b}_{i, 1}, \ldots, \bar{b}_{i, N}\big]^T \in \mathbb{R}^{(2 K + N) \times 1}$ is the $i$th particle. Moreover, $\mathbf{z}_{i}$ denotes the concatenation of the digital beamforming weights amplitude, phase, and variables corresponding to the PSs configurable phases, thus $N_{var} = 2 K + N$ is the number of variables. Note that PSO does not handle integer variables and since $b_{i,l}$ is an integer denoting the selected PS phase shift, we define $\bar{b}$ as a continuous variable in $[0, 1]$, and then, compute $b_{i,l}$ using $\bar{b}_{i,l}$ when evaluating the objective function.

\title{Particle Swarm Optimization (PSO) Procedure}
\date{}
\maketitle

\begin{algorithm}[t]
\caption{PSO-based waveform and Beamforming Design.}
\begin{algorithmic}[1]
\State \textbf{Inputs:} $w$, $c_1$, $c_2$, $I_{max}$, $N_p$, $\mathcal{K}$, $K$, $N_{var}$, $B$, ${P}_{dc}$, $X_{\max}$ 
\State \textbf{Outputs:} $P_c$, ${{X_{k}}, \phi_{k}, b_{i,l}}$
\State \textbf{Initialize:} 
\State $t = 1$, $v_{i,j} = 0$, $\mathbf{z}_{\min, j} = 0, \forall j$ 
\State $\mathbf{z}_{\text{max}, j} =X_{\max}, \quad \text{for } j\!=\!1, \ldots, K$
\State $\mathbf{z}_{\text{max}, j} = 2\pi, \quad \text{for } j\!=\!K+1, \ldots, 2K$
\State $\mathbf{z}_{\text{max}, j} = 1, \quad \text{for } j= 2K+1, \ldots, N_{\text{var}}$
\State Generate random swarm set $\mathcal{S}_{}$ with $\left|\mathcal{S}_{}\right|=N_p$ 
\For{$i = 1,..., N_p$}
    \State $b_{i,l} = \lfloor{\bar{b}_{i,l}(2^B - 1)}\rfloor, \forall l$
    \State $\mathbf{p}_{b_i}=\mathbf{z}_i$, $i^*=\operatorname{argmax}_i \xi\left(\mathbf{z}_i\right)$, $\mathbf{g}_{b}=\mathbf{z}_{i^*}$
\EndFor
\For{$t = 1,..., I_{\max}$}
    \For{$i = 1,..., N_p$}
    \For{$j = 1,..., N_{var}$}
      \State \hspace{-2mm}Choose $\mathcal{K} \subset \mathcal{S}_{}$ randomly
      \State \hspace{-2mm}Select ${r}_1$ and ${r}_2$ randomly in $(0,1)^{{N_{var}} \times 1}$
         \State \hspace{-12mm} ${v}_{{i},{j}}^{{t}}\!=\!w {v}_{{i}, {j}}^{{t}-1}\!+\!{c}_1 {r}_1\bigl(\mathbf {p}_{b_{{i}, {j}}}^{{t}-1}\!-\!\mathbf{z}_{{i}, {j}}^{{t}-1}\bigr)\!+\!{c}_2 {r}_2\bigl(\mathbf{g}_{b_{j}}^{t-1}\!-\!\mathbf{z}_{{i}, {j}}^{{t}-1}\bigr)$

        \State \hspace{-2mm}Update $\mathbf{z}_{i, j}^t=\mathbf{z}_{i, j}^{t-1}+v_{i, j}^{t-1}$, $b_{i,l} = \lfloor{\bar{b}_{i,l}(2^B - 1)}\rfloor$
    \hspace{-2mm}\EndFor
    \State \textbf{if} $\xi\left(\mathbf{z}_i\right)<\xi\left(\mathbf{p}_{b_i}\right)$ \textbf{then} $\mathbf{p}_{b_i} \leftarrow \mathbf{z}_i$ \textbf{end if}
      \State \textbf{if} $\xi(\mathbf{z}_i)<\xi\left(\mathbf{g}_{b}\right)$ \textbf{then} $\mathbf{g}_{b} \leftarrow \mathbf{z}_i$   \textbf{end if}
    \hspace{-2mm}\EndFor
\EndFor
\end{algorithmic}
\end{algorithm}
\begin{figure*}[!ht]
\centering
   \includegraphics[width=0.9\textwidth]{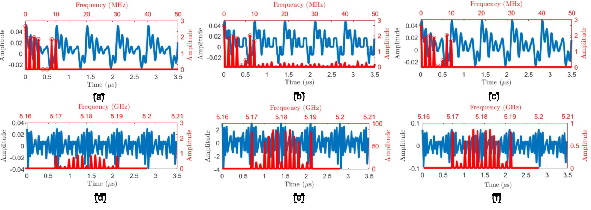}
   \vspace{-3mm}
    \caption{Multi-tone signal at different system blocks with $K$ = 8, $n_b$ = 2, $B$ = 3, $N$ = 25. (a) Digital multi-tone signal, (b) DAC output signal, (c) LPF output signal, (d) mixer output signal, (d) HPA output signal, and (e) RF signal at the ER.}
\label{fig_3}
\vspace{-1.5em}
\end{figure*}
Algorithm 1 illustrates the waveform and beamforming design algorithm, which requires several update-evaluate steps to reach a suboptimal solution. First, we generate an initial population of random particles and initialize the parameters. Next, for each particle, $\eqref{eq:44}$ is evaluated and $\mathbf{p}_{b_i}$ and $\mathbf{g}_{b}$ are obtained. Note that the variables corresponding to the phase shifts are continuous in the range $[0, 1]$; thus, this value should be up-scaled and converted to an integer to match the characteristic of the limited-resolution phase shifts, as in line 10. During the update steps (lines 18-19), the next velocity and the position of the particles are updated. Specifically, $v_{i, j}^t$ and $\mathbf{z}_{i, j}^t$ are the velocity and the position of the $j$th dimension of the $i$th particle at iteration $t$, respectively. After that, for each particle, the fitness function is calculated, while the $\mathbf{p}_{b_i}$ of each particle and the $\mathbf{g}_{b}$ of the swarm are updated (line 19). The procedure is repeated until the maximum number of iterations, i.e., $I_{max}$, is reached. Notably, the complexity of the algorithm increases with the number of variables, thus, more iterations and swarm size may be needed to attain a proper solution.\par{}
PSO has several parameters that can be modified to improve its performance depending on the optimization problem. For instance, ${w}$ is the inertia weight, which is a balancing factor between exploration and exploitation. Large $w$ offers more exploration, while smaller $w$ facilitates more exploitation. Moreover, $c_1$ and $c_2$ are the weights for personal experience and social interaction. Additionally, $\mathbf{z}_{\min }$ and $\mathbf{z}_{\max }$ are $N_{v a r}$-dimensional vectors denoting the lower and upper bound of the optimization variables, respectively. 
\vspace{-1mm}
\subsection{Complexity Analysis}
\vspace{-1mm}
  The complexity of Algorithm 1 is determined by the computation complexity of \eqref{eq:44} and the intricacy of the PSO. The bottleneck of this process is the computation time of $\bar{v}_{\text{out}}$, which scales with $K$, $N$, the number of time samples in the charging period, i.e., $N_{tp}$, and IDFT computational complexity, i.e, $\mathcal{O}\left(K \log K\right)$, as seen in \eqref{eq:1} and \eqref{eq:16}. Therefore, the complexity of calculating $\bar{v}_{\text{out}}$ is $\mathcal{O}\left(K^2  N N_{tp}\log K\right)$. Since the process is repeated for all particles, the total complexity of the utilized PSO algorithm is $ \mathcal{O}\left(N_p K^2 N N_{tp}\log K\right)$.
\section{Numerical Analysis}
We consider the ER is located at 3~m distance in the boresight direction of the UPA with an EH requirement of ${P}_{d c}=20\ \mu \mathrm{W}$ \cite{clerckx2018beneficial}. The array is square-shaped with $V$ = $H$ = $\sqrt{N}$, while we set $X_{\max} = 300$ V, $BW = 10$~MHz, $\tilde{f}_s = 100$~MHz, and $f_{ca} = 5.18$~GHz unless otherwise stated \cite{zhang2023waveform}. The rest of the parameters are listed in Table~I.

\begin{figure}
    \centering
    \includegraphics[width=0.7\columnwidth]{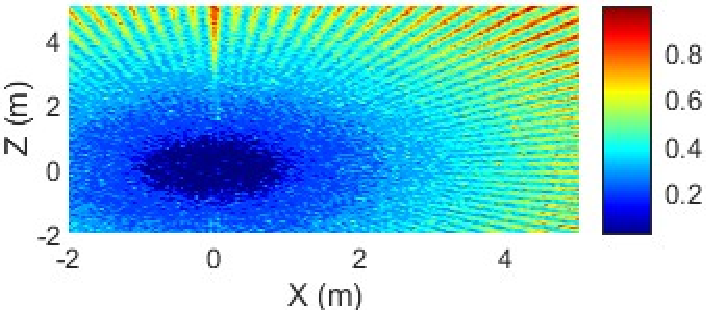}
    \vspace{-2mm}
    \caption{The normalized consumed power with respect to the path loss for different ER positions in the area.}
    \label{fig_61}
    \vspace{-4mm}
\end{figure}

\setlength{\tabcolsep}{6pt}
\renewcommand{\arraystretch}{0.9}
\begin{table}[!t]
\vspace{-0.5mm}
\caption{System model parameters.}
\begin{tabular}{|c |c |c |c |c |c|}
\toprule
\textbf{Symbol} & \textbf{Value} & \textbf{Ref}  & \textbf{Symbol} & \textbf{Value} & \textbf{Ref}    \\ \midrule
$V_{dd}$ & 3 V & \cite{cui2005energy}  & $A_s$ & 10      &  \cite{zhang2023waveform}                            \\ 
$R_s$ & 50 $\Omega$    & \cite{moghadam2017waveform} & $\beta$ & 4      & \cite{zhang2023waveform}    \\
$L_{ps}$ & 0.5 dB      & \cite{eisenbeis2019comparison}  & $R_L$ & 1.6 K$\Omega$  &  \cite{moghadam2017waveform}  \\
$P_{mix}$ & 23 mW & \cite{lin2016energy} & $A$ & 1 V & \cite{cui2005energy} \\ 
$I_0$ & 5 $u$A & \cite{moghadam2017waveform}  &
$P_{lo}$ & 5 mW & \cite{lin2016energy} \\ 
$V_0$ & 25.86 mV & \cite{moghadam2017waveform} &
$G$ & 10 & \cite{joung2014survey} \\ 
$\alpha$ & 1 & \cite{cui2005energy} &
$\eta$ & 1.05 & \cite{moghadam2017waveform} \\ 
$C_p$ & 1 pF & \cite{cui2005energy} &
$I$ & 10 $u$A & \cite{cui2005energy} \\ 
 \bottomrule 
\end{tabular}
\footnotesize
\noindent {For simplicity, we have used $R_{in} = R_{out} = 1 \, \Omega$. In practice, these values would depend on the design of the HPA.}
\label{text neck exp details}
\vspace{-1.5em}
\end{table}
Fig.~\ref{fig_3} illustrates the multi-tone signal waveform in time and frequency domains after passing through system blocks. In Fig.~\ref{fig_3} (a) and (b), the effect of low-resolution DAC on the signal can be seen, which generates high out-of-band frequencies. Moreover, it can be observed in Fig.~\ref{fig_3} (c)-(e) that the frequency response above $BW$ in baseband and outside $[f_{ca}-BW, f_{ca} + BW]$ is zero. Fig.~\ref{fig_61} showcases the normalized $P_c$ in the area. As expected, the $P_c$ increases as we move toward further distances from the transmitter, while it is seen that some spatial directions are more power-consuming. 

\begin{figure}
    \centering
    \includegraphics[width=0.8\columnwidth]{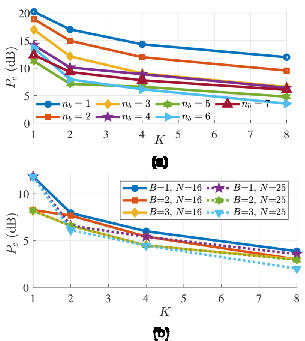}
    \vspace{-2mm}
    \caption{Power consumption versus $K$ for (a) different $n_b$ values (b) different $B$ and $N$.}
    \label{fig_5}
    \vspace{-6mm}
\end{figure}

\begin{figure}[t!]
    \centering
    \includegraphics[width=0.8\columnwidth]{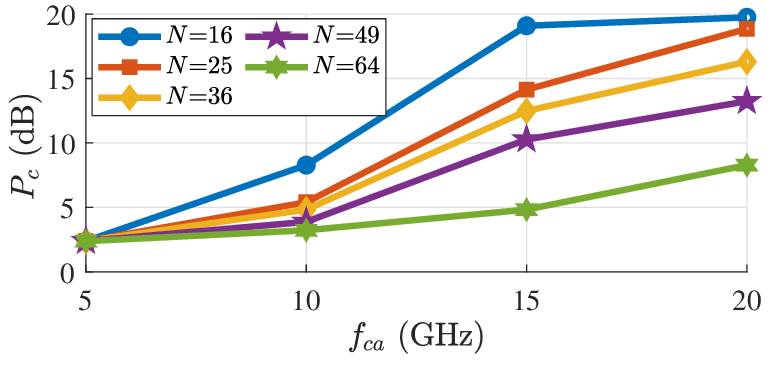}
    \vspace{-3mm}
    \caption{Power consumption versus the operating frequency for different $N$ values, $K = 8$, $n_b = 3$, and $B = 3$.}
    \label{fig_6}
    \vspace{-8mm}
\end{figure}

Fig.~\ref{fig_5} (a) showcases $P_c$ as a function of $K$ for different DAC resolutions. It is seen that increasing $n_b$ leads to reducing $P_c$, which is due to better waveform and beamforming capability. Note that at some point using more bits, e.g., $n_b \ge 7$, $P_{dac}$ drastically increases while the waveform/beamforming gains become marginal, which leads to an increase in overall system $P_c$. Fig.~\ref{fig_5} (b) shows $P_c$ versus $K$ for different $B$ and $N$. It can be observed that the performance is improved by using higher $B$ for PSs, which allows for more control over the direction and shape of the transmitted signal, leading to achieving more harvested power with less power consumption. Finally, Fig.~\ref{fig_6} illustrates $P_c$ increasing with $f_{ca}$ for different $N$. It can be also observed that increasing $N$ leads to reducing $P_c$. Moreover, the efficiency of HPA tends to decrease at higher frequencies. This is because HPA faces greater challenges in maintaining linearity due to increased parasitic effects, leading to more required input power to achieve the same output power levels.
\vspace{-5mm}
\section{Conclusion}
\vspace{-1mm}
In this paper, we considered a single-user multi-antenna analog RF-WPT system. We first formulated a joint waveform and beamforming optimization aiming to minimize the system power consumption while meeting EH requirements. Then, we proposed an optimization approach relying on PSO to solve the problem. The results evinced that the power consumption decreases as the DAC/PSs resolution and antenna length increase, while increasing frequency leads to larger power consumption. In future studies, we may investigate the performance of the system in multi-user scenarios and the impact of digital/hybrid transmit antenna architectures. 

\bibliographystyle{IEEEtran}
\bibliography{main}
\vspace{-2mm}
\end{document}